\documentclass{aa}
\usepackage{graphicx}
\usepackage{graphics}
\usepackage{epsfig}

\newcommand{\be}{\begin{eqnarray}}
\newcommand{\ee}{\end{eqnarray}}

\begin{document}

\title{A model of the light curves of Gamma-Ray Bursts}
\author{W. H. Lei, D. X. Wang, B. P. Gong, and C. Y. Huang}

\titlerunning{A model of the light curves of GRBs}
\authorrunning{Lei et al.}

\institute{Department of Physics, Huazhong University of Science and
Technology, Wuhan, 430074, China}

\offprints{D. X. Wang\\ \email{dxwang@mail.hust.edu.cn}}
\date{Received 11 August 2006 / Accepted 8 March 2007}

\abstract{}{An extreme Kerr black hole (BH) surrounded by a
precessing disk is invoked to explain the light curves of gamma-ray
bursts (GRBs) based on the coexistence of the Blandford-Znajek (BZ)
and the magnetic coupling (MC) processes.} {The overall shape of the
light curves and the duration of GRBs are interpreted by the
evolution of the half-opening angle of the magnetic flux on the BH
horizon, and the complex temporal structures are modulated by the
precession and nutation of the jet powered by the BZ process.} {The
time profile of the emission exhibits a fast rise and a slow decay
due to the effect of the evolution of the half-opening angle. The
light curves of several GRBs are well fitted by this model with only
six free parameters.}{}


\maketitle

\section{Introduction}

Gamma ray bursts (GRBs) are possibly the most luminous objects in the
Universe. Extremely high energy released in very short timescales suggests
that GRBs involve the formation of a black hole (BH) via a catastrophic
stellar collapse event or possibly a neutron star merger, implying that an
inner engine could be built on an accreting BH (Piran 2004).

Among a variety of mechanisms for powering GRBs, the BZ process
(Blandford {\&} Znajek 1977) has the unique advantage of providing
``clean'' (free of baryonic contamination) energy by extracting
rotating energy from a BH and transferring it in the form of
Poynting flow in the outgoing energy flux (Lee et al. 2000).
Recently, observations and theoretical considerations have linked
long-duration GRBs with ultrabright Type Ib/c SNe (Galama et al.
1998; Bloom et al. 1999). Brown et al. (2000) proposed a specific
scenario for a GRB-SN connection. They argued that the GRB is
powered by the BZ process and the SN is powered by the MC process,
which is regarded as one of the variants of the BZ process (van
Putten 1999; Blandford 1999; Li 2000, 2002; Wang et al. 2002).
However, they failed to distinguish the fractions of the energy for
these two objects. More recently, van Putten and collaborators
proposed a dominant spin-connection between the central BH and
surrounding high-density matter (van Putten 2001; van Putten {\&}
Levinson 2003). It is based on similar shapes in topology of the
torus magnetosphere with the magnetosphere of pulsars, when viewed
in a poloidal cross-section. This description points towards
complete calorimetry on GRB-SNe, upon including an unexpectedly
powerful long-duration burst of gravitational-radiation,
representing most of the spin-energy liberated from the central BH.

Lei et al. (2005, hereafter LWM05) proposed a scenario for GRBs in
Type Ib/c SNe, invoking the coexistence of the BZ and MC processes.
In LWM05 the half-opening angle of the magnetic flux tube on the
horizon is determined by the mapping relation between the angular
coordinate on the BH horizon and the radial coordinate on the
surrounding accretion disk. In this scenario the half-opening angle
evolves to zero with the spinning-down BH. This effect shuts off a
GRB, and the overall timescale of the GRB could be fitted by the
lifetime of the open magnetic flux on the horizon.

Besides the feature of high energy released in very short
durations, most GRBs are characterized by highly variable light
curves with complex temporal behavior. The usual explanations of
the temporal structures range from multiple shock fronts running
into the ambient medium (Sari, Narayan {\&} Piran 1996), expanding
shells with brighter patches and dimmer regions (Fenimore, Madras
{\&} Nayakshin 1996), to repeated series of pulses with Gaussian
or power-law profiles (Norries et al. 1996). However, a clear
physical explanation is lacking. Some authors (Fargion {\&} Salis
1996; Blackman, Yi {\&} Field 1996; Fargion 1999; Portegies Zwart,
Lee {\&} Lee 1999, hereafter PZLL99; Portegies Zwart {\&} Totani
2001, hereafter PZT01; Fargion {\&} Grossi 2006) suggested that
the light curves of GRBs can be explained by a beamed emission
from a precessing jet. PZLL99 constructed ad hoc a function $I(t)$
to describe the outline of the light curves, which is
characterized by three timescales: a fast rise with timescale
$\tau _{rise} $, a plateau phase with timescale $\tau _{plat} $,
and a stiff decay with timescale $\tau _{decay} $. The complex
temporal structures are modulated by the precession and nutation
of the jet. Although their model successfully fits several light
curves of GRBs, the origin of the function $I(t)$ is lacking.

In fact, the evolution of the half-opening angle on the BH horizon described
in LWM05 provides a natural explanation for the outline of the light curves.
Based on LWM05 and PZLL99, in this paper, we intend to combine the evolution
of the half-opening angle with the precession and nutation of the jet to fit
the complex light curves of GRBs.

The paper is organized as follows. In section 2 we derive the jet
luminosity per steradian by the mapping relation between angular
coordinate on the BH horizon and the cylindrical radius of the
collimated jet. Based on the evolution of an extreme Kerr BH we
obtain an intrinsic time variation of the gamma-ray emission. It is
shown that the time profile of the emission exhibits a fast rise and
a slow decay. In section 3, the light curves of several GRBs are
fitted by combining the evolution of the half-opening angle with the
precession and nutation of the jet. In section 4, we summarize the
main results and discuss some issues related to our model.
Throughout this paper the geometry units $G=c=$1 are used.

\section{Time profile of jet luminosity}

The poloidal configuration of the magnetic field is shown in Fig. 1,
which is adapted from van Putten (2001).

%-----------------------------------------------------------------
\begin{figure}
\centering
%\resizebox{\hsize}{!}{\includegraphics[]{fig1.eps}}
%\includegraphics[trim = 0mm 6mm 0mm 7mm, clip, width=8cm,angle=0]{fig1.eps}
%\centerline{\includegraphics[width=3.00in,height=3.00in]{fig1.eps}}
\centerline{\includegraphics[width=7cm]{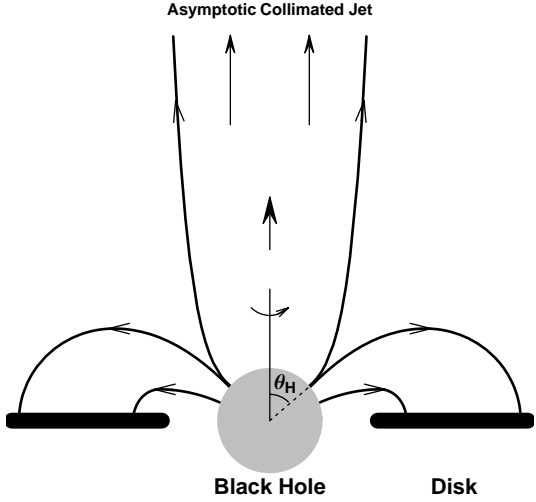}}
%\centerline{\includegraphics[width=80mm,height=80mm]{fig1.eps}}
\caption[]{The polodial topology of the magnetosphere of a disk
surrounding a rapidly rotating BH (not to scale). Along the open
magnetic flux tube, an asymptotic collimated jet may be formed.}
\label{fig1}
\end{figure}
%-----------------------------------------------------------------

In Fig. 1 the angle $\theta _H $ is the half-opening angle of the
open magnetic flux tube, indicating the angular boundary between
open and closed field lines on the horizon. The angle $\theta _H $
can be determined by (Wang et al., 2003)

\begin{equation}
\label{eq1}
\cos \theta _H = \int_1^\infty {\mbox{G}\left( {a_ * ;\xi ,n} \right)d\xi }
.
\end{equation}

Eq. (\ref{eq1}) is derived based on the conservation of magnetic
flux with three assumptions: (i) the magnetic field on the horizon
$B_H $ is constant, (ii) the magnetic field on the disk surface $B_D
$ varies as a power law with the radial coordinate of the disk, and
(iii) the magnetic flux connecting the BH with its surrounding disk
takes precedence over that connecting the BH to the remote load. In
Eq. (\ref{eq1}) $a_\ast \equiv J / M^2$ is the BH spin defined in
terms of the BH mass $M$ and the angular momentum $J$, the parameter
$n$ is the power-law index for the variation of $B_D $, i.e., $B_D^
\propto \xi ^{ - n}$, and $\xi \equiv r / r_{ms} $ is the radial
coordinate on the disk, which is defined in terms of the radius
$r_{ms} \equiv M\chi _{ms}^2 $ of the marginally stable orbit
(Novikov {\&} Thorne 1973). The function $G(a_\ast ;\xi ,n)$ is
given by

\be
G\left( {a_ * ;\xi ,n} \right)&=& \frac{\xi ^{1 - n}\chi _{ms}^2
}{2}\sqrt {\frac{1 + a_ * ^2 \chi _{ms}^{ - 4} \xi ^{ - 2} + 2a_ *
^2 \chi _{ms}^{ - 6} \xi ^{ - 3}}{1 + a_ * ^2 \chi _{ms}^{ - 4} +
2a_ * ^2 \chi _{ms}^{ - 6} }} \times  \nonumber\\ & & \frac{1}{\sqrt
{\left( {1 - 2\chi _{ms}^{ - 2} \xi ^{ - 1} + a_
* ^2 \chi _{ms}^{ - 4} \xi ^{ - 2}} \right)} } .
\label{eq2}
\ee

By using Eq. (\ref{eq1}) we have the curves of $\theta _H $ versus
$a_\ast $ with different values of the power-law index $n$ as shown
in Fig. 2, in which the half-opening angle $\theta _H $ always
increases monotonically with increasing $n$ for a given BH spin
$a_\ast $.

%-----------------------------------------------------------------
\begin{figure}
\centering
%\resizebox{\hsize}{!}{\includegraphics[]{fig2.eps}}
%\includegraphics[trim = 0mm 6mm 0mm 7mm, clip, width=8cm,angle=0]{fig2.eps}
\centerline{\includegraphics[width=70mm,height=70mm]{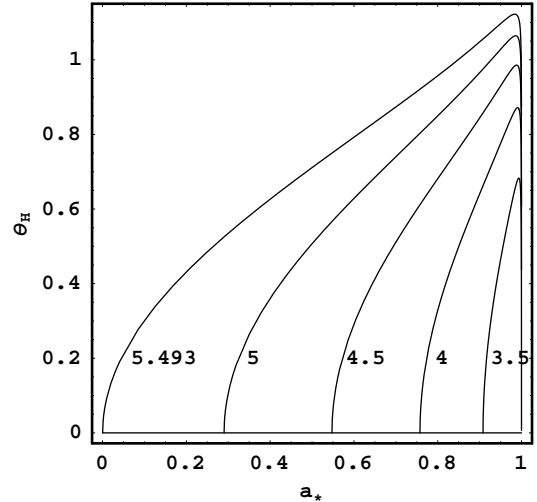}}
\caption[]{The curves of $\theta _H $ vs $a_ * $ for $n$=3.5, 4,
4.5, 5, 5.493.} \label{fig2}
\end{figure}
%-----------------------------------------------------------------

A very interesting feature shown in Fig. 2 is that the angle $\theta
_H $ evolves non-monotonically with decreasing $a_ * $, i.e., it
increases very rapidly as $a_ * $ spins down from unity and then
decreases slowly to zero for $3.003 \le n \le 5.493$. Therefore, the
time-scale of the duration of the burst is given by the lifetime of
rapid spin of the BH as mentioned by van Putten (2001). This
evolution characteristic can be applied to fitting GRBs in two
aspects: (\ref{eq1}) determining the duration of the GRBs as
described in LWM05, and (\ref{eq2}) shaping the time profile of jet
luminosity. To see the latter effect, we will derive the jet
luminosity in \S 2.1, and investigate its time profile in \S 2.2.

\subsection{Jet luminosity}

As is well known, the rotating energy is extracted from a BH in the
form of Poynting flow in the BZ process, and this energy dissipates
and accelerates electrons to produce gamma-rays through synchrotron
radiation or inverse Compton scattering. Although some works
approach the magnetic dissipation in jets, its origin remains
unclear (Lyutikov {\&} Blandford 2003; Spruit, Daigne, {\&}
Drenkhahn 2001). As a simple analysis, we introduce a parameter
$\varepsilon _\gamma $ to denote the fraction of BZ energy converted
into gamma-ray energy.

The BZ power transferred through two adjacent magnetic surfaces between
$\theta $ and $\theta + d\theta $ on the BH horizon is given by (Wang et al.
2002)

\begin{equation}
\label{eq3}
d\tilde {P}_{BZ} = dP_{BZ} / P_0 = 2a_\ast ^2 \frac{k(1 - k)\sin ^3\theta
}{2 - (1 - q)\sin ^2\theta }d\theta ,
\end{equation}

\noindent
where $P_0 \equiv B_H^2 M^2$, $q \equiv \sqrt {1 - a_\ast ^2 } $, and $k$ is
the ratio of the angular velocity of open magnetic field lines to that of
the BH. For the optimal BZ power, $k = 0.5$ is taken.

There are a variety of approaches to studying collimating jet flows,
such as the curvature force exerted by the toroidal field (Sakurai
1985), the pressure of the poloidal magnetic field surrounding the
jet (Blandford 1993; Spruit 1994), and the pressure of the
baryon-rich wind supported by the surrounding disk (van Putten 2001;
van Putten {\&} Levinson 2003). The collimation of the flow is
determined by the cross-field force-balance in the direction
perpendicular to the field. Fendt (1997, hereafter F97) obtained the
numerical solution of the stream equation for this force-balance of
a force-free magnetic jet. The asymptotic jet can be collimated into
a cylindrical shape with a jet radius of several light cylinder
radii.

In order to work out an analytical model we adopt F97's scenario and assume
that the poloidal component of the magnetic field of the collimated jet,
$B_L $, varies as $B_L / B_H = e^{ - \lambda x}$, where $x \equiv R / r_H $
is the cylindrical radius of the collimated jet in terms of the radius of
the BH horizon. Based on the conservation of magnetic flux, we have

\begin{equation}
\label{eq4}
\Delta \Psi _H = B_H^ 2\pi \left( {\varpi \rho } \right)_{r = r_H } d\theta
= B_L^ 2\pi RdR.
\end{equation}
Substituting $\left( {\varpi \rho } \right)_{r = r_H } = \sum _{r =
r_H } \sin \theta = 2Mr_H \sin \theta $ into Eq. (\ref{eq4}), we
have

\begin{equation}
\label{eq5}
{2\sin \theta d\theta } \mathord{\left/ {\vphantom {{2\sin \theta d\theta }
{\left( {1 + q} \right)}}} \right. \kern-\nulldelimiterspace} {\left( {1 +
q} \right)} = xe^{ - \lambda x}dx.
\end{equation}
Integrating Eq. (\ref{eq5}), we obtain the mapping relation between
the angular coordinate $\theta $ on the BH horizon and the
cylindrical radius $x$ the collimated jet as follows

\begin{equation}
\label{eq6} \cos \theta = 1 - \frac{1 + q}{2\lambda ^2}\left[ {1 -
(1 + \lambda x)e^{ - \lambda x}} \right].
\end{equation}

Defining $R_L \equiv c / \Omega _F = 2 / \Omega _H $ as the light
cylindrical radius of the collimated jet, we have $x_L \equiv R_L /
r_H = 4 \mathord{\left/ {\vphantom {4 {a_\ast }}} \right.
\kern-\nulldelimiterspace} {a_\ast }$. Following F97, the radius
$R_{jet} $ of a cylindrically collimated jet should be less than
$4R_L $. In this work, we assume $R_{jet} \approx R_L $, and the
field lines from $\theta _H $ near the BH will connect to $R_L $ in
the collimated region as shown in Fig. 1. Thus the relation between
$\theta _H $ and $x_L $ is expressed by

\begin{equation}
\label{eq7}
\cos \theta _H = 1 - \frac{1 + q}{2\lambda ^2}\left[ {1 - (1 + \lambda x_L
)e^{ - \lambda x_L }} \right].
\end{equation}

Following PZLL99, the angle dependence of the jet is given
approximately as $x = x_f \sin \psi $, where $x_f \equiv {r_f }
\mathord{\left/ {\vphantom {{r_f } {r_H }}} \right.
\kern-\nulldelimiterspace} {r_H }$ is the freezeout distance from
the BH to the remote load, where the photons can leave freely. The
opening angle $\psi _{jet} $ is fixed by $\sin \psi _{jet} \approx
x_L / x_f $ as PZLL99, and a small opening angle $\psi _{jet} =
6{\degr}$ with $x_f = 10x_L $ is assumed.

Based on the above equations we express the jet luminosity per steradian for
an observer at angle $\psi $ with respect to the central locus of the jet,
i.e.,

\be
\label{eq8}
\tilde {L}_{jet} (a_\ast ,\theta _H ,\psi ) \equiv
L_{jet} / P_0 &=& \varepsilon _\gamma d\tilde {P}_{BZ} / d\Omega
\nonumber \\ &=& \frac{\varepsilon _\gamma d\tilde {P}_{BZ} }{2\pi
\sin \psi d\psi } \nonumber \\ &=& \frac{\varepsilon _\gamma }{2\pi
\sin \psi }\frac{d\tilde {P}_{BZ} }{d\theta }\frac{d\theta
}{dx}\frac{dx}{d\psi },
\ee

\noindent
where $\theta _H = \theta _H (a_\ast ,n)$ is determined by
Eq. (\ref{eq1}).

\begin{figure}[htbp]
\centerline{\includegraphics[width=70mm,height=70mm]{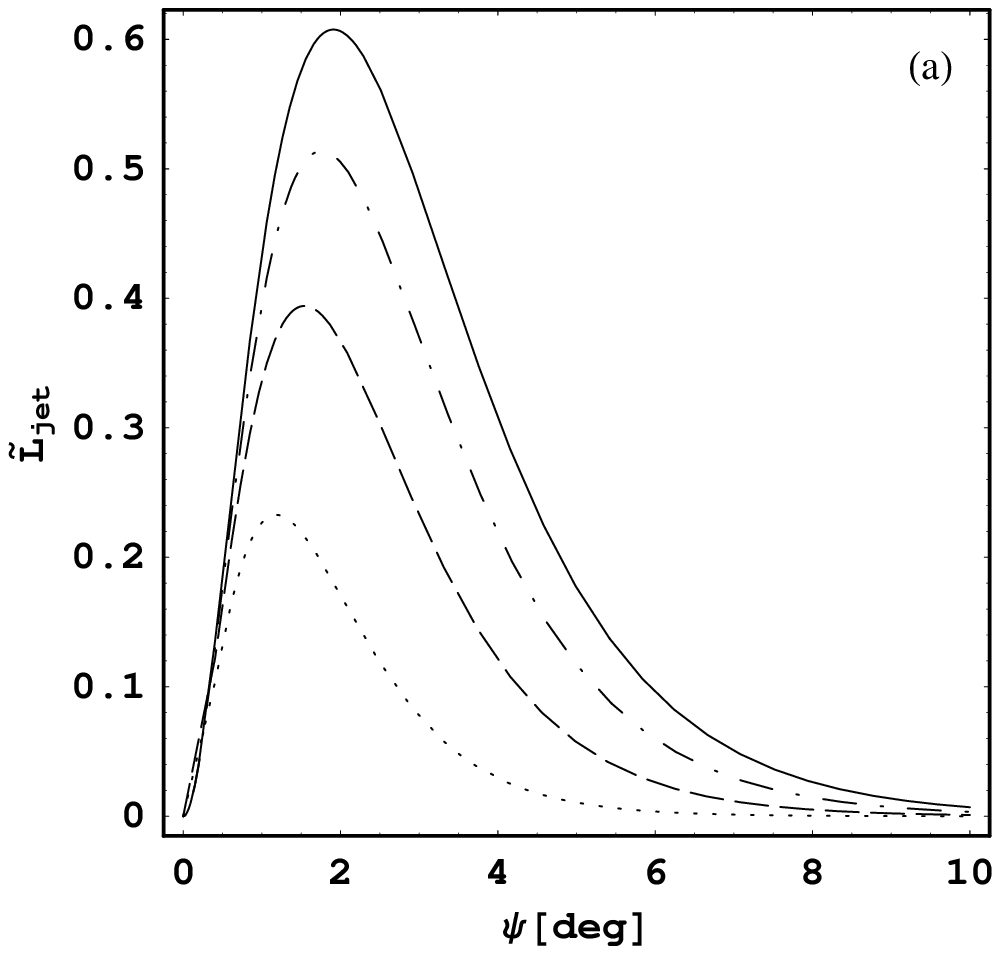}}
\centerline{\includegraphics[width=70mm,height=70mm]{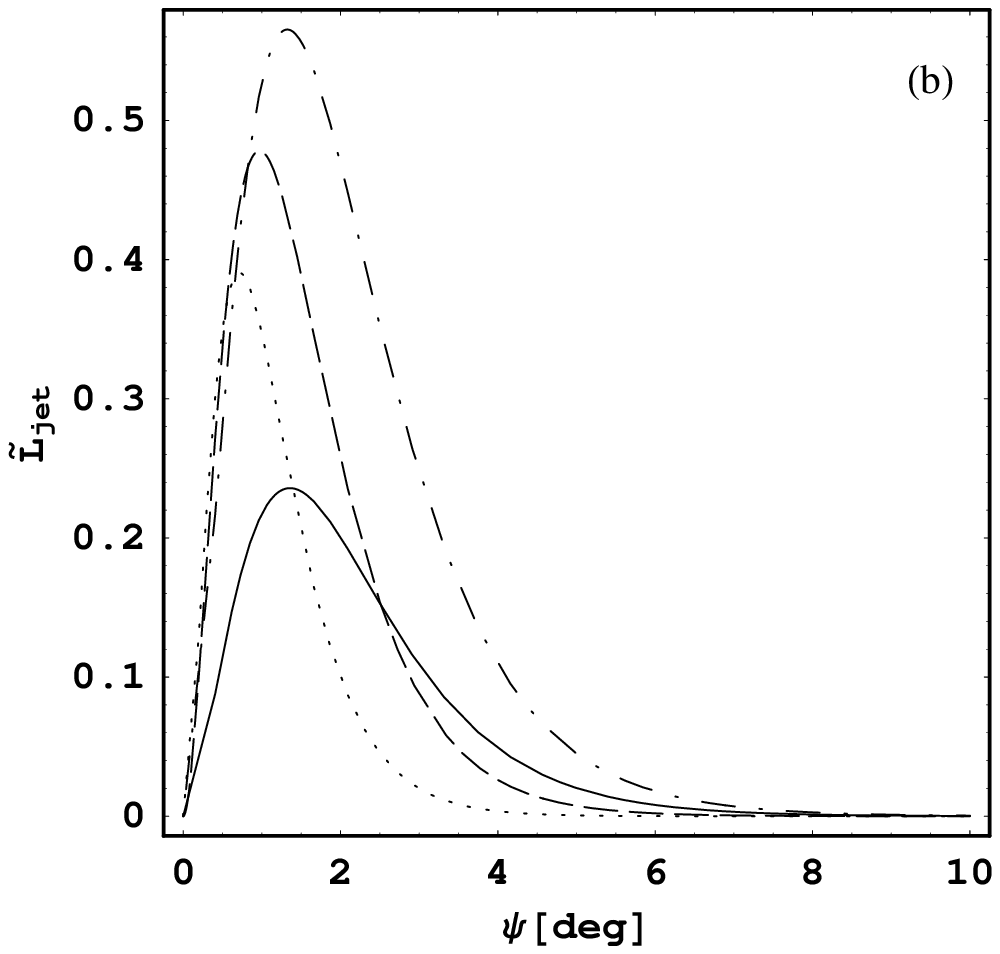}}
\caption[]{$\psi $-dependence of $\tilde {L}_{jet} $ a) for $a_\ast
= 0.99$ with $n = $5, 4.5, 4 and 3.5 in solid, dash-dotted, dashed
and dotted lines, respectively, and b) for $n = 5$ with $a_\ast =
0.9999$, 0.9, 0.8 and 0.7 in solid, dash-dotted, dashed and dotted
lines, respectively.} \label{fig3}
\end{figure}

Taking $\varepsilon _\gamma \simeq 0.15$ in calculations as given by
van Putten et al. (2004), we have $\tilde {L}_{jet} $ varying with
$\psi $ as shown in Fig. 3. From Fig. 3 we find that the value of
$\tilde {L}_{jet} $ is sensitive to $n$ and $a_\ast $. It vanishes
at $\psi = 0$, and attains its peak value near the spin axis of the
BH. In addition, we find that both the value and the angular
coordinate of the peak $\tilde {L}_{jet} $ strongly depend on the
values of $a_\ast $ and $n$, while the variation of $\tilde
{L}_{jet} $ with $\psi $ remains similar.

The $\theta _H $ dependence of $\tilde {L}_{jet} $ for the given
values of $a_\ast $ and $\psi $ is depicted in Fig. 4, and we find
that the value of $\tilde {L}_{jet} $ is sensitive to that of
$\theta _H $.

\begin{figure}[htbp]
\centerline{\includegraphics[width=70mm,height=70mm]{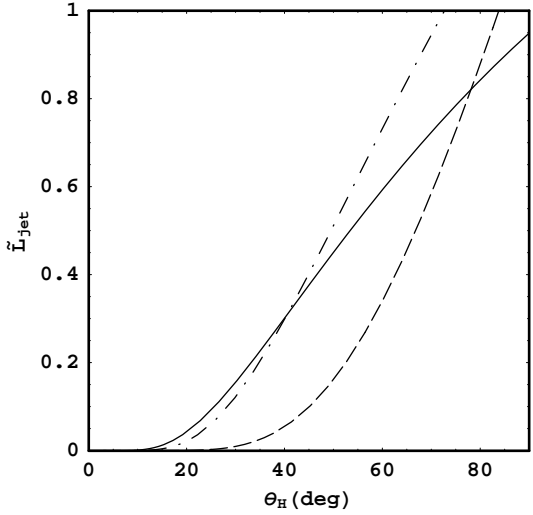}}
\caption[]{The curves of $\tilde {L}_{jet} $ vs $\theta _H $ for
$a_\ast = 0.9$ and $\psi = 1{\degr}$(solid line), $a_\ast = 0.7$ and
$\psi = 1{\degr}$(dash-dotted line), and $a_\ast = 0.9$ and $\psi =
3{\degr}$(dashed line).}
 \label{fig4}
\end{figure}

Combining the evolution of $\theta _H $ with the BH spin, we have
the curves of the luminosity $\tilde {L}_{jet} $ versus $a_\ast $ as
shown in Fig. 5.

\begin{figure}[htbp]
\centerline{\includegraphics[width=70mm,height=70mm]{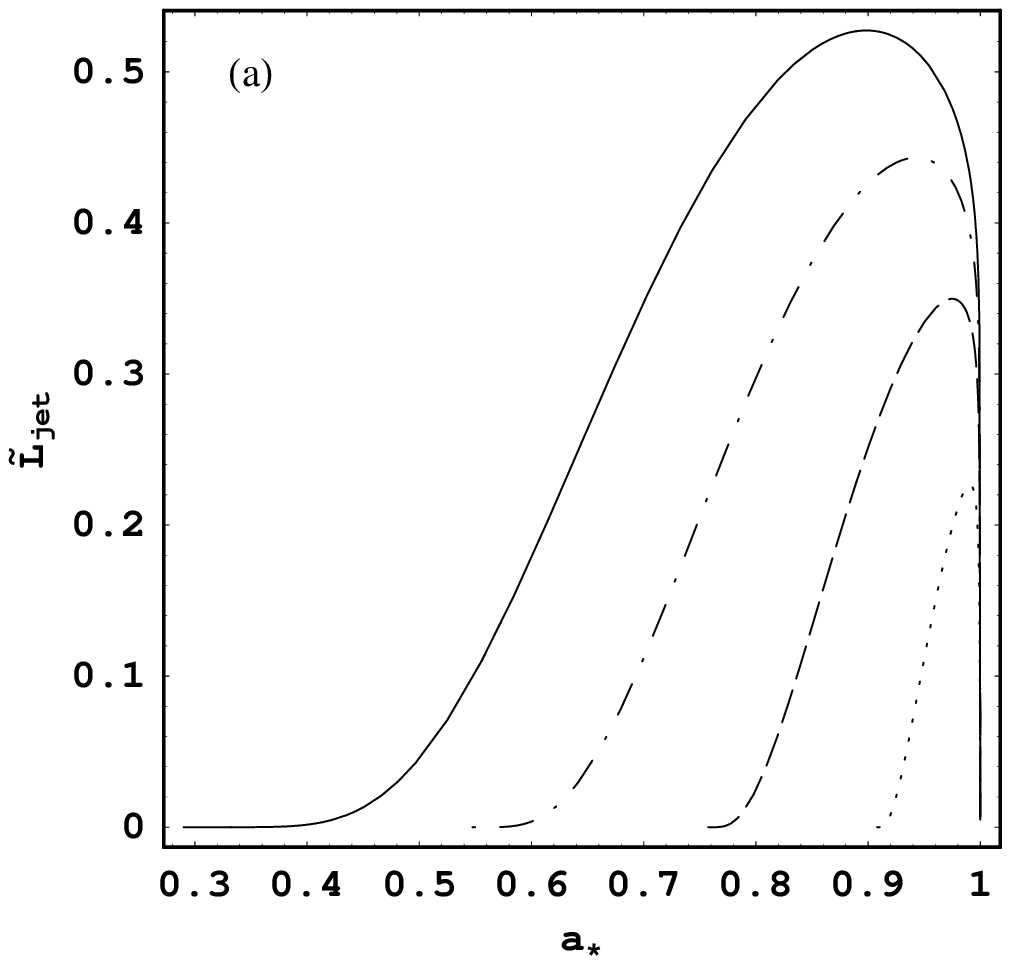}}
\centerline{\includegraphics[width=70mm,height=70mm]{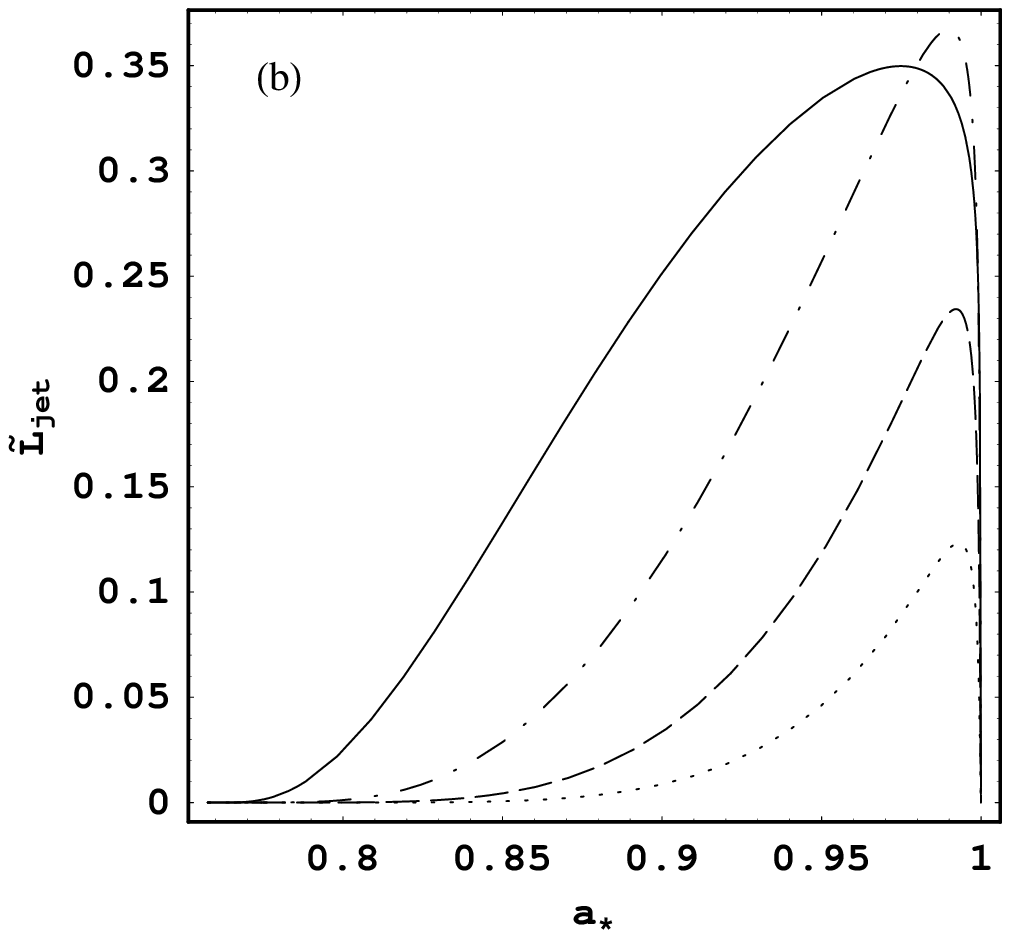}}
\caption[]{The curves of $\tilde {L}_{jet} $ vs $a_\ast $ a) for
$\psi = 1{\degr}$ with $n = $5, 4.5, 4 and 3.5 in solid,
dash-dotted, dashed and dotted lines, respectively; b) for $n = 4$
with $\psi = 1{\degr}$, $2{\degr}$, $3{\degr}$ and $4{\degr}$ in
solid, dash-dotted, dashed and dotted lines, respectively.}
 \label{fig5}
\end{figure}

Inspecting Fig. 5, we find that the luminosity $\tilde {L}_{jet} $
evolves non-monotonically with the spin of an extreme Kerr BH,
increasing very fast at first and then decreasing slowly. This
result is understandable by considering the non-monotonic evolution
of $\theta _H $ with $a_ * $ and the monotonic evolution of $L_{jet}
$ with $\theta _H $ as shown in Figs. 2 and 4, respectively.

\subsection{Intrinsic time dependence of jet luminosity}

Based on the BH evolution we can derive the time profile of $L_{jet} $.
Considering that the angular momentum is transferred from the rapidly
rotating BH to the disk, on which a positive torque is exerted, we think
that the accretion onto the BH is probably halted. This state is discussed
as the suspend accretion state by van Putten {\&} Ostriker (2001), and as
the nonaccretion solution by Li (2002). In this state, the BH evolution is
governed by the BZ and MC processes. Based on the conservation of energy and
angular momentum we have the evolution equations of the BH as follows,

\begin{equation}
\label{eq9}
{dM} \mathord{\left/ {\vphantom {{dM} {dt}}} \right.
\kern-\nulldelimiterspace} {dt} = - (P_{BZ} + P_{MC} ),
\end{equation}

\be
\label{eq10} {da_ * } \mathord{\left/ {\vphantom {{da_ * }
{dt}}} \right. \kern-\nulldelimiterspace} {dt} &=& - M^{ - 2}(T_{BZ}
+ T_{MC} ) + 2M^{ - 1}a_
* (P_{BZ} + P_{MC} ) \nonumber \\ &=& B_H^2 MA(a_\ast ,n),
\ee

\noindent
and

\begin{equation}
\label{eq11}
A(a_\ast ,n) \equiv - (T_{BZ} + T_{MC} ) / T_0 + 2a_ * (P_{BZ} + P_{MC} ) /
P_0 .
\end{equation}

The powers and torques for the BZ and MC processes are expressed as (Wang et
al. 2003),

\begin{equation}
\label{eq12}
\tilde {P}_{BZ} \equiv {P_{BZ} } \mathord{\left/ {\vphantom {{P_{BZ} } {P_0
}}} \right. \kern-\nulldelimiterspace} {P_0 } = 2a_ * ^2 \int_0^{\theta _H }
{\frac{k\left( {1 - k} \right)\sin ^3\theta d\theta }{2 - \left( {1 - q}
\right)\sin ^2\theta }}
\end{equation}

\begin{equation}
\label{eq13}
{\tilde {T}_{BZ} \equiv T_{BZ} } \mathord{\left/ {\vphantom {{\tilde
{T}_{BZ} \equiv T_{BZ} } {T_0 }}} \right. \kern-\nulldelimiterspace} {T_0 }
= 4a_ * \left( {1 + q} \right)\int_0^{\theta _H } {\frac{\left( {1 - k}
\right)\sin ^3\theta d\theta }{2 - \left( {1 - q} \right)\sin ^2\theta }}
\end{equation}

\begin{equation}
\label{eq14}
{\tilde {P}_{MC} \equiv P_{MC} } \mathord{\left/ {\vphantom {{\tilde
{P}_{MC} \equiv P_{MC} } {P_0 }}} \right. \kern-\nulldelimiterspace} {P_0 }
= 2a_ * ^2 \int_{\theta _H }^{\pi / 2} {\frac{\beta \left( {1 - \beta }
\right)\sin ^3\theta d\theta }{2 - \left( {1 - q} \right)\sin ^2\theta }}
\end{equation}

\begin{equation}
\label{eq15}
{\tilde {T}_{MC} \equiv T_{MC} } \mathord{\left/ {\vphantom {{\tilde
{T}_{MC} \equiv T_{MC} } {T_0 }}} \right. \kern-\nulldelimiterspace} {T_0 }
= 4a_ * \left( {1 + q} \right)\int_{\theta _H }^{\pi / 2} {\frac{\left( {1 -
\beta } \right)\sin ^3\theta d\theta }{2 - \left( {1 - q} \right)\sin
^2\theta }}
\end{equation}

\noindent
where $T_0 \equiv B_H^2 M^3$, and $\beta \equiv \Omega _D / \Omega _H $ is
the ratio of the angular velocity of the disk to that of the BH.

Substituting Eqs. (\ref{eq9}) and (\ref{eq10}) into Eq. (\ref{eq8}),
we have the curves of $L_{jet} (t)$ varying with time $t$ for
different values of $\psi $ and $n$ as shown in Fig. 6. The magnetic
field on the BH horizon $B_H = 10^{15}G$, the initial BH mass $M(0)
= 7M_ \odot $ and the initial BH spin $a_\ast (0) = 1$ are assumed
in calculations.

\begin{figure}[htbp]
\centerline{\includegraphics[width=70mm,height=70mm]{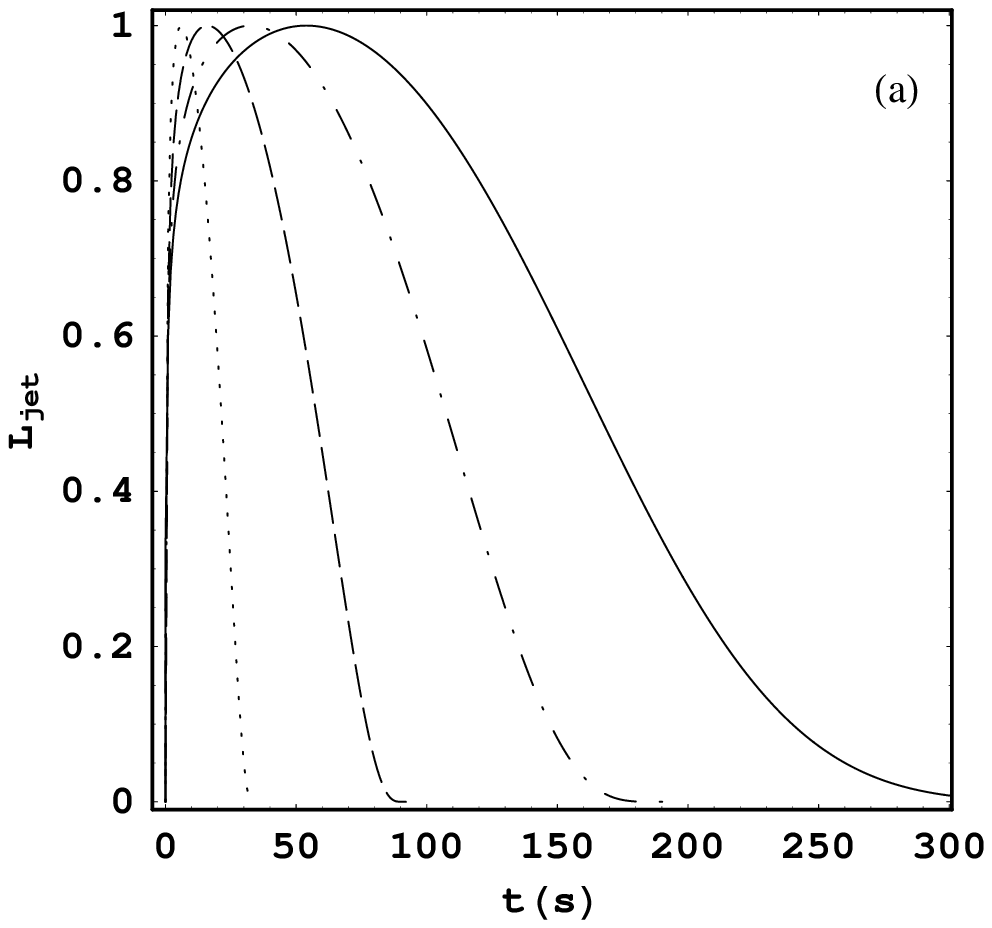}}
\centerline{\includegraphics[width=70mm,height=70mm]{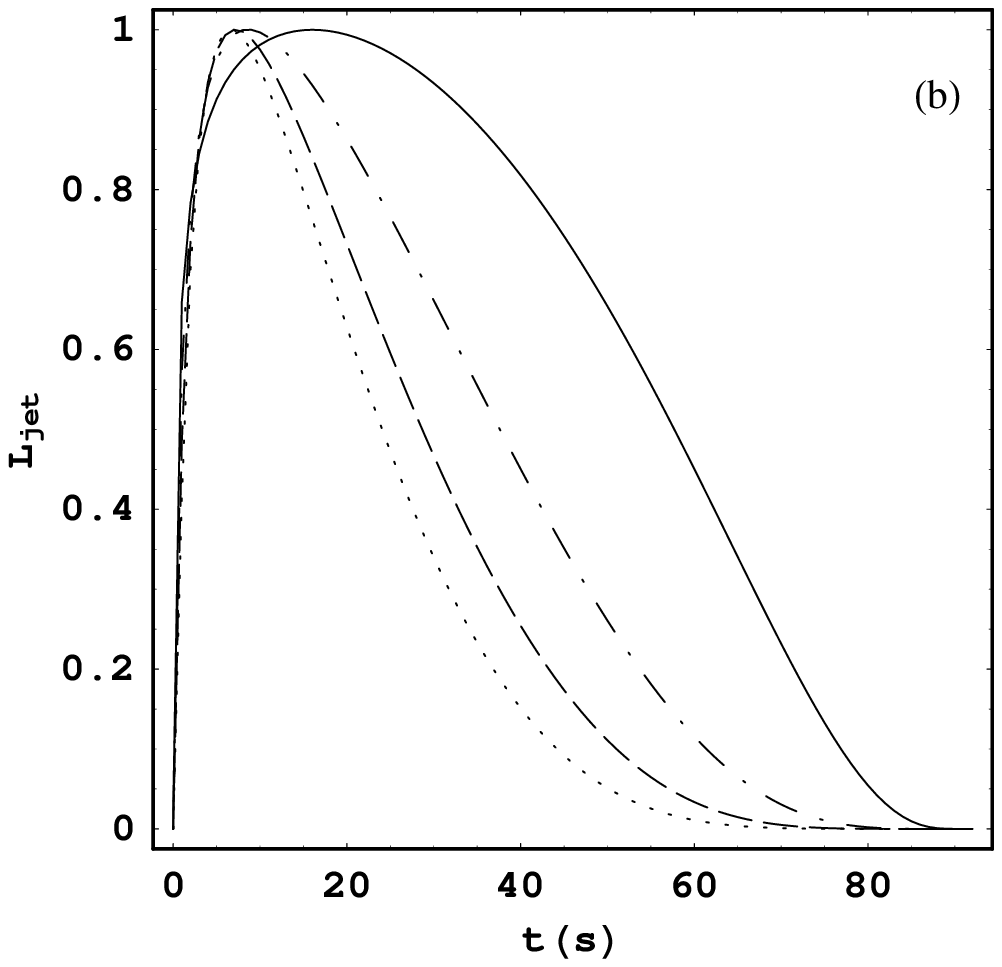}}
\caption[]{The curves of $L_{jet} $ vs time $t$ for different $n$
and observed angle $\psi $, each curve is normalized to unity at its
maximum: a) for $\psi = 1{\degr}$ with $n = $5, 4.5, 4 and 3.5 in
solid, dash-dotted, dashed and dotted lines, respectively; b) for $n
= 4$ with $\psi = 1{\degr}$, $2{\degr}$, $3{\degr}$ and $4{\degr}$
in solid, dash-dotted, dashed and dotted lines, respectively.}
\label{fig6}
\end{figure}

As shown in Fig. 6, the time profiles of $L_{jet} $ exhibit a
feature of fast-rise and slow-decay, which arises from the
non-monotonic evolution characteristic of $\theta _H $ in terms of
the BH spin as shown in Fig. 2. Thus the intrinsic time profile of
the jet luminosity is derived naturally in our model rather than
assumed ad hoc, and it resembles that given in PZLL99. The shapes of
the profiles can be adjusted by changing the values of $n$ and $\psi
$. It has been argued in LWM05 that the durations of various GRBs
can be fitted by changing the value of $n$ for the given initial BH
spin.

\section{Fitting observed light curves}

\subsection{Precessing jet}

The misalignment in the spin axis of the BH and the angular momentum
axis of the binary causes the accretion disk around the BH to
precess, which is known as the slaved disk precession. PZLL99
investigated the precession of the accretion disk in the gamma-ray
binary, in which a neutron star fills its Roche lobe and transfers
its mass to a BH. It is shown that this system has a precession
period of about a second. The Lense-Thirring precession appears for
a Kerr BH, if the accretion disk is inclined with respect to the
equatorial plane of the BH (Lense {\&} Thirring 1918). Recently,
Reynoso et al. (2006, hereafter RRS06) studied the possible effect
of Lense-Thirring precession on neutrino-cooled accretion disks. It
is found that the precession period can be much less than 1 second.

As the magnetic field is assumed to be anchored in the disk, this
would result in the jet perpendicular to the midplane of the disk,
and the precession and nutation of the disk directly implies the
precession and nutation of the jet. To describe the jet
kinematically we adopt the angular evolution of the jet in terms of
the spherical angles $\theta _{jet} $ and $\phi _{jet} $ given in
PZLL99 as follows,

\begin{equation}
\label{eq16}
\phi _{jet} = \Omega _{pre} (t - t_0 ) + \frac{\Omega _{pre} }{\Omega _{nu}
}\sin (\Omega _{nu} t),
\end{equation}

\begin{equation}
\label{eq17}
\theta _{jet} = \theta _{jet}^0 + \frac{\Omega _{pre} }{\Omega _{nu} }\tan
\theta _{jet}^0 \cos (\Omega _{nu} t),
\end{equation}

\noindent where $\Omega _{pre} = 2\pi / \tau _{pre} $ and $\Omega
_{nu} = \pm 2\pi / \tau _{nu} $ are the angular frequencies of
precession and nutation with periods $\tau _{pre} $ and $\tau _{nu}
$, respectively. The direction of the observer can be assigned as
$\theta _{obs} $ and $\phi _{obs} $. A schematic picture of the
precessing disk and jet is shown in Fig. 7, and the angle between
the observer and the central locus of the jet is

\begin{equation}
\label{eq18}
\psi = \cos ^{ - 1}(\hat {r}_{obs} \cdot \hat {r}_{jet} )
\end{equation}

\begin{figure}[ht]
% \psfig{figure=M3radios.bmp, width=8.0cm}
%\vspace{-0.5cm}
%\includegraphics{f2.eps}
\centering
%                         l   b     r    t
\includegraphics[trim = 10mm 160mm 100mm 13mm, clip, width=7cm]{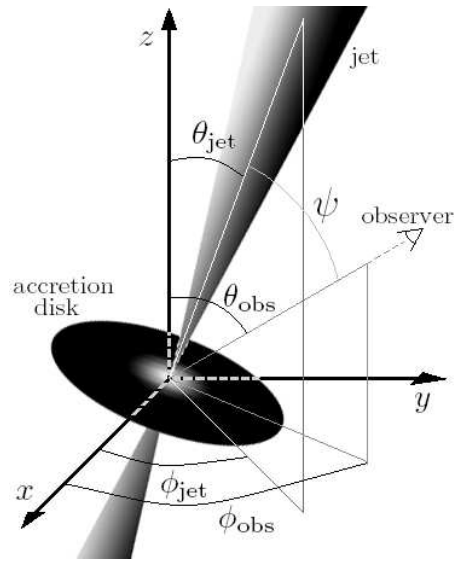}
%\centerline{\includegraphics[width=6in height=6in]{fig7.ps}}
%\includegraphics[clip, width=3cm]{f2.eps}
%\psfig{figure=f2.eps}
%\vspace{-16cm}
\caption{A schematic picture of a precessing disk with jet (adapted
from RRS06).} \label{fig7}
\end{figure}

\subsection{Fittings}

Substituting Eq. (\ref{eq18}) into Eq. (\ref{eq8}), we obtain the
gamma-ray luminosity of a prcessing jet $L_{jet} (t)$, which is
adopted to fit the observed complex light curves of GRBs. The model
contains six free parameters to fit the light curves (as described
in Table 1), and the fitting procedures are described as follows.

\begin{table}
\caption{Free parameters that may vary per burst} \label{table:1}
\centering
\begin{tabular}{c c} \hline \hline
Parameters & Note \\
\hline
$n$& The power-law index for the variation \\
& of the magnetic field on the disc \\
$\tau _{pre} $& Precession period \\
$R_\Omega \equiv \Omega _{nu} / \Omega _{pre} $&
Frequency ratio of nutation to precession \\
$\theta _{jet}^0 $&
Initial angular coordinate of a jet \\
$\theta _{obs} ,\phi _{obs} $&
Observation angles \\
\hline
\end{tabular}
\end{table}

The first step is to determine the background. Following PZLL99,
this is done by averaging the count rate of the initial, $\sim $1800
data. The background is subtracted in our fitting process.

Next, we introduce the start time $\tau _{start} $ and the end time
$\tau _{end} $, which are the times of the start-up and shut-off of
the real burst, respectively. As a rough estimate, we take $\tau
_{start} $ and $\tau _{end} $ as the first and last time to reach
about 1{\%} of the peak count rate. Our model fits the light curve
from $\tau _{start} $ to $\tau _{end} $.

From Figs. 2 and 6 (a), we find that the duration of GRB is
sensitive to the value of $n$. Therefore, the first free parameter
$n$ is chosen to satisfy the observed duration ($\tau _{end} - \tau
_{start} )$ in our model. Substituting $n$ into Eqs. (\ref{eq10})
and (\ref{eq11}), we obtain the evolution functions of BH mass
$M(t)$, BH spin $a_\ast (t)$ and half-opening angle $\theta _H (t)$.
The precessing effect is described by the function $\psi (t)$.
Substituting these functions into Eq. (\ref{eq8}), we can produce
the time variability of the observed flux.

Finally, we use the simulated annealing to fit the observed bursts. The
other five parameters in Table 1, i.e.,$\tau _{pre} $, $R_\Omega $, $\theta
_{jet}^0 $, $\theta _{obs} $ and $\phi _{obs} $ are chosen freely in this
simulation. After each iteration we determine the $\chi ^2$ from the fitting
of $L_{jet} (t)$ to the observed burst profile. This value is minimized by
the annealing algorithm.

We apply this model to fit several observed GRBs by adjusting the
value of the above six free parameters (see Table 1). The data of
the light curves are taken from BATSE and HETE. We fit the light
curves of GRBs from the third energy channel, i.e., 100-300keV for
BATSE, and 30-400keV for HETE. The fittings results are presented in
Fig. 8 with the parameters listed in Table 2.

Inspecting Fig. 8, we find that the light curves of these GRBs can
be fitted by our model. Some simple light curves of GRBs are fitted
in a rather satisfactory way as shown in Figs. 8a, 8c, 8e and 8g,
and complicated light curves are fitted with good overall shapes as
shown in Figs. 8b and 8d. Fig. 8f shows the fitting for the more
complicated burst GRB 030519a for which most of the peaks are well
fitted except for the third one. GRB 920701 is a peculiar asymmetric
burst (Romero et al. 1999; RRS06), with a slower rise and a faster
decay compared with the other bursts. It turns out that this type of
burst can be also fitted by our model as shown in Fig. 8b.

\section{Discussion}

In this paper we discuss a model for fitting the light curves of
GRBs based on the coexistence of the BZ and MC processes, combining
the evolution of an extreme Kerr BH surrounded by a precessing disk.
Comparing with PZLL99 , PZT01 and RRS06, our model can explain both
the intrinsic time profile and the observed duration of GRB. The
fittings are done with only six parameters. However, there are still
several issues related to this model and the fittings.

First, a precessing jet tends to produce light curves in which
subsequent peaks appear at periodic intervals. However, the observed
gamma-ray bursts show no evidence for periodicities at any
time-scale. PZT01 solved this problem by considering the interaction
between the jet and the interstellar medium. In our model, the
distance from the BH to the emission region is fixed. If this
distance varies during the GRB, the asymmetry of the subsequent
peaks in the light curves could be explained by our model.

Second, the individual peaks produced by precession clearly lack the
observed strong asymmetries, i.e., a fast rise and a slow decay
(Fenimore, Madras {\&} Nayakshin 1996). In our model, the time
variability of the central engine is rather smooth. Thus our model
cannot explain the strong asymmetries. As described in PZT01, the
effect of the curvature of the jet front or the cooling of gamma-ray
emitting electrons can make the pulse profile a `fast rise and slow
decay'. We will take these effects into account in future work.

Third, this model is good in fitting the overall shapes of the light
curves of the bursts, but it is difficult to fit the smaller details
such as the variability on time scales of milliseconds. Recently,
Wang et al. (2006) discussed the modulations of the screw
instability in the BZ process on the light curves of GRBs, and they
argued that the variability timescales of ten milliseconds can be
interpreted by two successive flares due to the screw instability of
the magnetic field. It is also shown that the individual peaks
produced by releasing and recovering magnetic energy have a fast
rise and slow decay profile. We shall improve the fitting of the
light curves of GRBs by combining the jet precession with screw
instability of the magnetic field in future work.

We cannot perform satisfactory fits for all kinds of GRB light
curves with this simple model, especially for those displaying
extraordinary periods of quiescence between bursts. Fig. 8f is one
of these cases, where a time-lag in the third peak is not
reproduced. This type of burst is hard to fit by the
jet-time-profile as given in Fig. 6, i.e., the feature of fast-rise
and slow-decay. The combinations of parameters listed in Table 2 are
not unique in the fittings. For example, the light curve of GRB
920701 can be fitted as well as Fig. 8b with the parameters
($n$=3.10; $\tau _{pre} $=5.18, $R_\Omega $=8.36; $\theta _{jet}^0
$=1.02, $\theta _{obs} $=1.19, $\phi _{obs} $=242.58).

\begin{acknowledgements}
We thank the anonymous referee for numerous constructive
suggestions. This work is supported by the National Natural Science
Foundation of China under Grant Numbers 10573006 and 10121503. This
research has made use of the HETE and BATSE data.
\end{acknowledgements}

{}

\newpage

\begin{figure*}
   \centering
   \includegraphics[width=6cm]{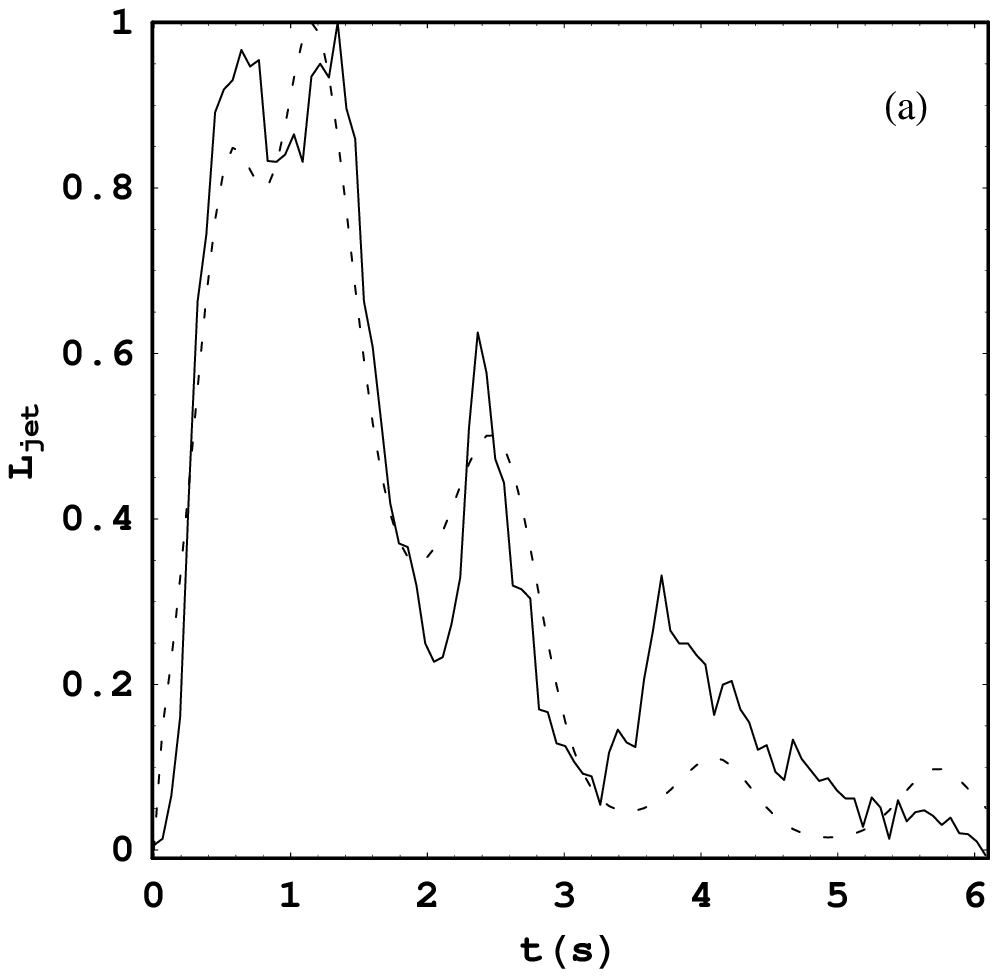}
   \includegraphics[width=6cm]{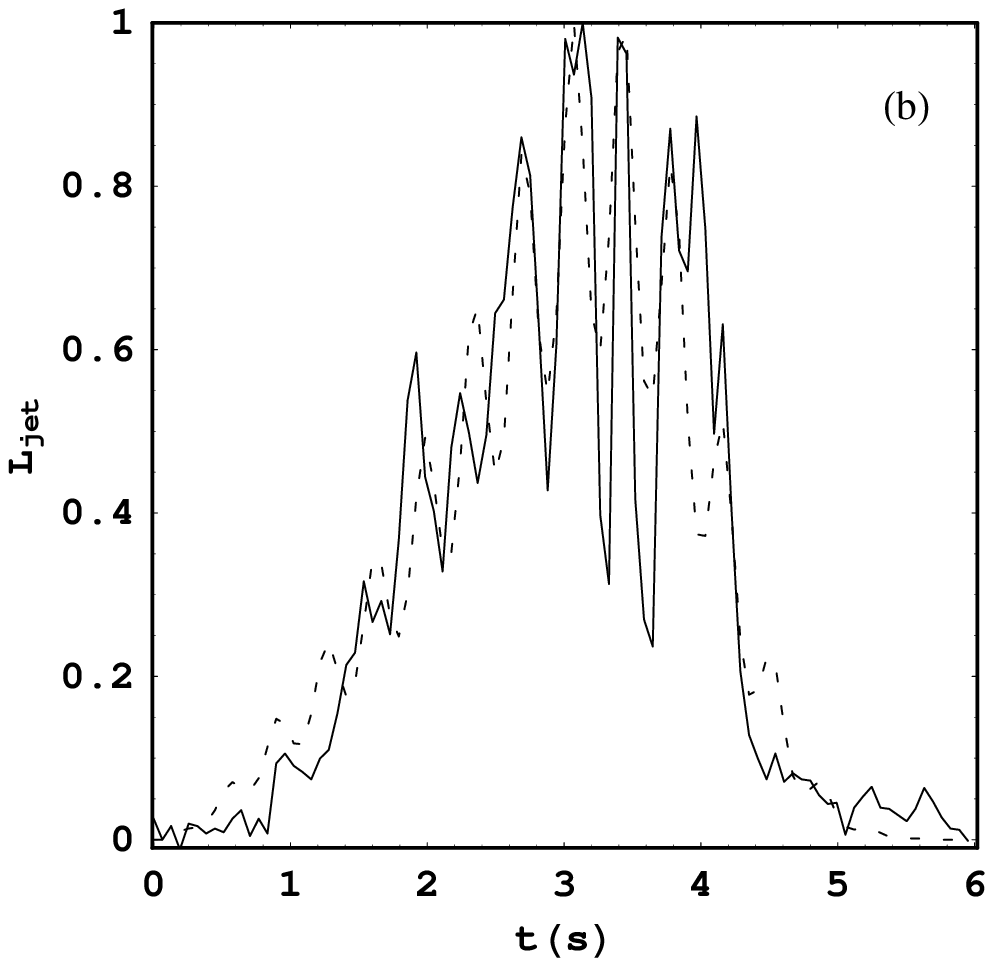}
   \includegraphics[width=6cm]{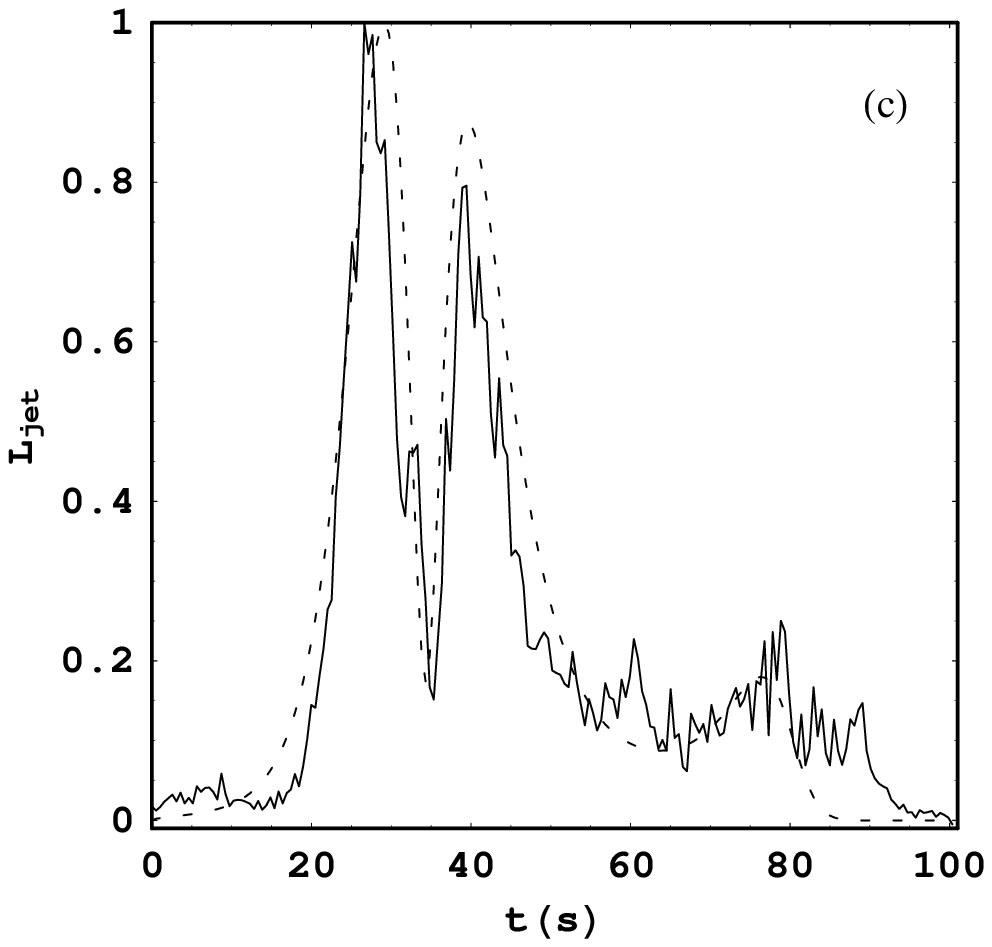}
   \includegraphics[width=6cm]{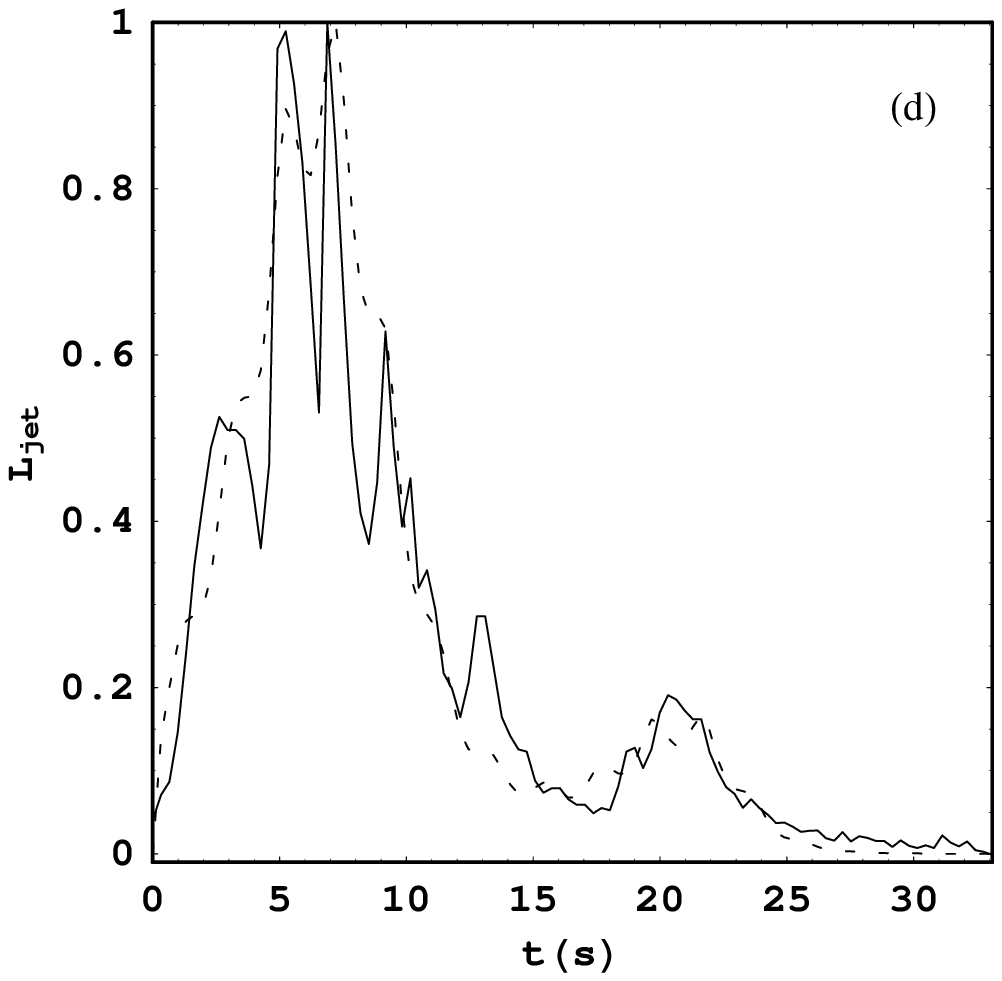}
   \includegraphics[width=6cm]{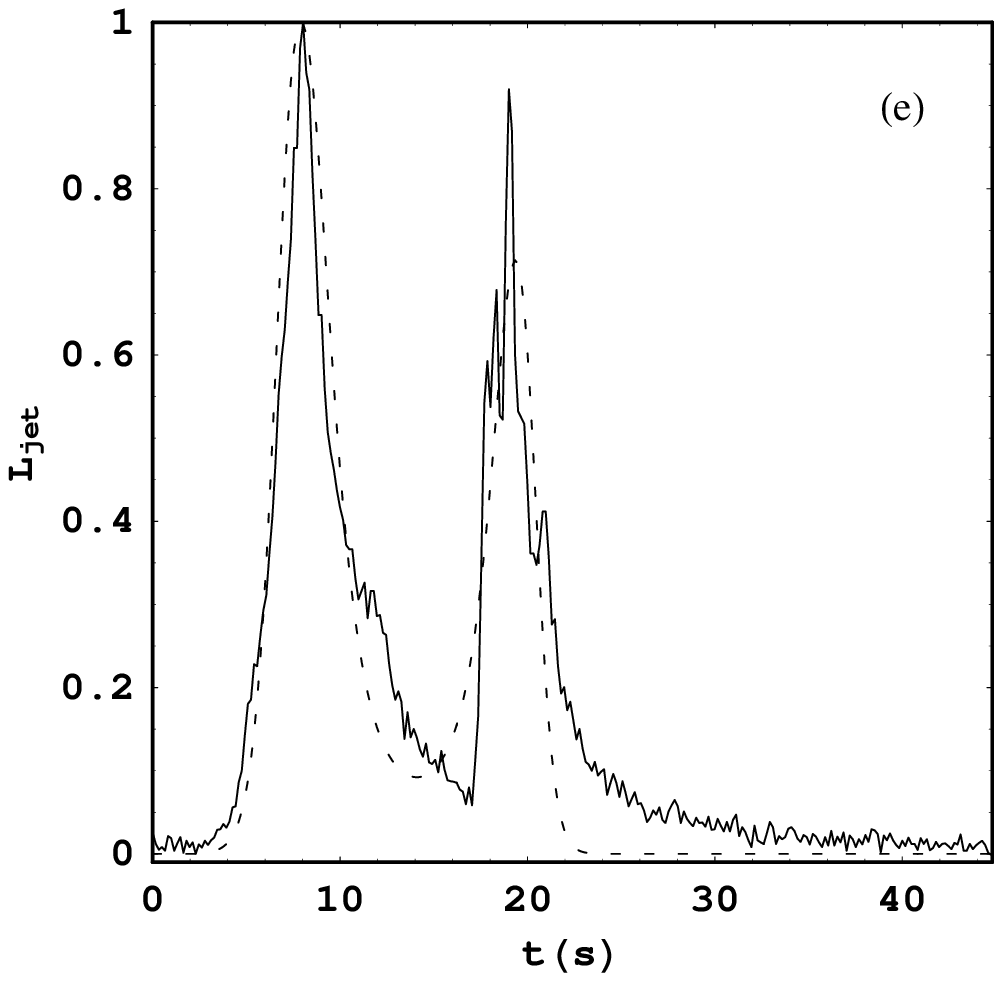}
   \includegraphics[width=6cm]{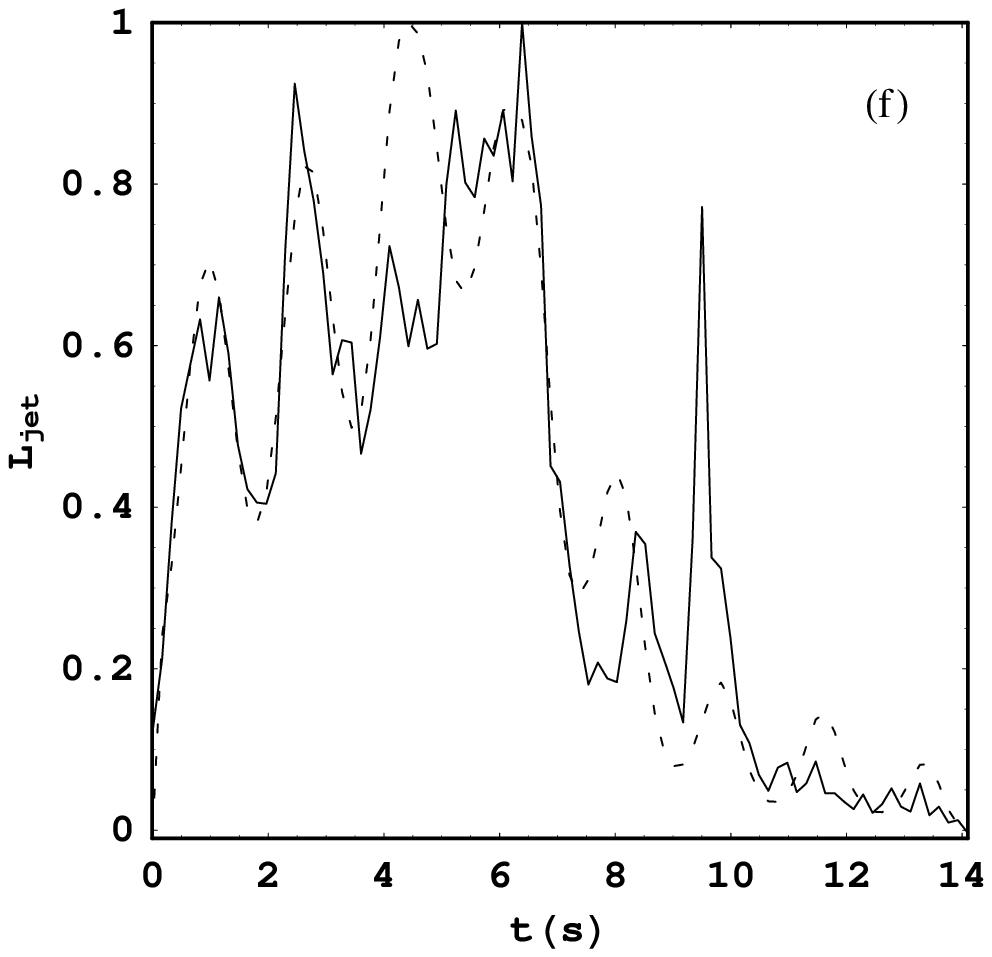}
   \includegraphics[width=6cm]{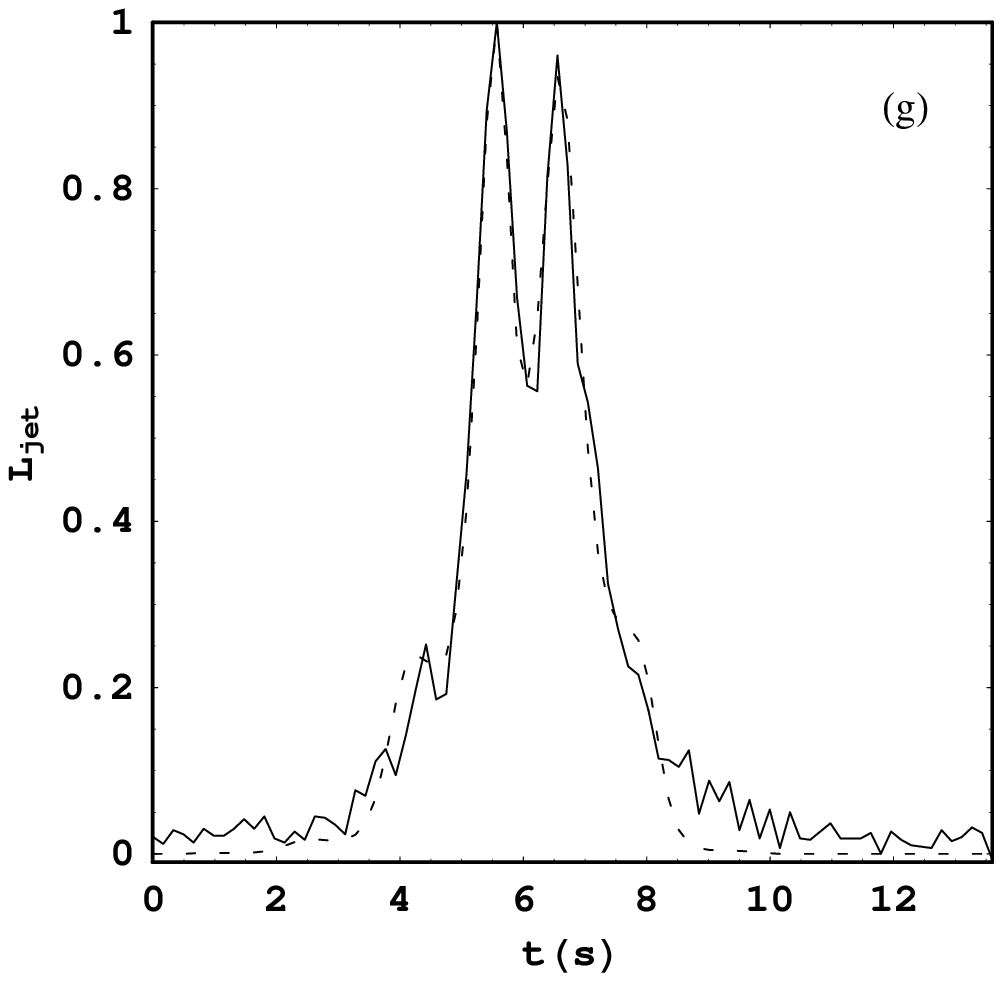}
\caption{Fits of the light curves from the third energy channel of
(a) GRB 910717, (b) GRB 920701, (c) GRB 990123, (d) GRB 001225, (e)
GRB 030329, (f) GRB 030519a and (g) GRB 031111a. The real bursts
profile and the fitting curves are plotted in full and broken
curves, respectively.} \label{fig8}
\end{figure*}

\begin{table*}
\caption{Values of the parameters for fitting several GRBs}
\label{table:2} \centering
\begin{tabular}{c c c c c c c c c c} \hline \hline
GRB & $\tau _{start} $ \par (s) & $\tau _{end} $ \par (s) & $n$&
$\tau _{pre} $ \par (s) & $R_\Omega $ & $\theta _{jet}^0 $ \par
(deg) & $\theta _{obs} $ \par (deg) & $\phi _{obs} $ \par (deg) &
$\chi ^2$  \\
\hline
910717& 1.86& 7.94& 3.15& 7.00& 4.22& 1.78& 0.89& 39.16& 0.88 \\
920701& 1.02& 6.98& 3.10& 7.24& 20.01& 2.63& 0.69& 188.25& 1.52 \\
990123& 0.51& 100.86& 4.10& 128.37& 1.11& 3.89& 2.97& 149.17& 2.05 \\
001225& 0.28& 33.37& 3.60& 14.68& 6.08& 0.74& 4.28& 172.58& 0.84 \\
030329& 7.24& 51.99& 3.70& 70.89& 2.60& 11.16& 11.20& 69.88& 1.78 \\
030519a& 0.09& 14.18& 3.26& 7.80& 4.35& 1.54& 0.41& 259.86& 1.63 \\
031111a& 2.56& 16.16& 3.26& 22.13& 12.86& 5.58& 4.96& 101.63& 0.14 \\
\hline
\end{tabular}
\label{tab2}
\end{table*}


\begin{thebibliography}{}
\bibitem{}Blackman, E. G., Yi, I., {\&} Field, G. B. 1996, ApJ, 473, L79
\bibitem{}Blandford, R. D., {\&} Znajek, R. L. 1977, MNRAS, 179, 433
\bibitem{}Blandford, R. D. 1999, in ASP Conf. Ser. 160, Astrophysical Discs: An EC Summer
School, ed. J. A. Sellwood {\&} J. Goodman (San Francisco: ASP), 265
\bibitem{}Blandford, R. D. 1993, in D. Burgarella, M. Livio and C. O'Dea, eds.,
Astrophysical Jets, Cambridge: Cambridge University Press, 15.
\bibitem{}Bloom, J. S., et al. 1999, Nature, 401, 453
\bibitem{}Brown, G. E., Lee, C.-H., Wijers, R. A. M. J., Lee, H. K., Israelian, G.,
{\&} Bethe H. A. 2000, NewA, 5, 191
\bibitem{}Camenzind, M. 1987, A{\&}A, 184, 341
\bibitem{}Fargion, D., {\&} Salis, A. 1996, astro-ph/9605166
\bibitem{}Fargion, D. 1999, A{\&}AS, 138, 507
\bibitem{}Fargion, D., {\&} Grossi, M. 2006, ChJAA, 6S1, 342
\bibitem{}Fendt, C. 1997, A{\&}A, 319, 1025 (F97)
\bibitem{}Fenimore E. E., Madras, C.D., {\&} Nayakshin, S. 1996, ApJ, 473, 998
\bibitem{}Galama, T. J., et al. 1998, Nature, 395, 670
\bibitem{}Lee, H. K., Wijers, R. A. M. J., {\&} Brown, G. E. 2000, Phys. Rep.,
325, 83
\bibitem{}Lei, W. H., Wang, D. X., {\&} Ma, R. Y. 2005, ApJ, 619, 420 (LWM05)
\bibitem{}Lense, J., {\&} Thirring, H. 1918, Phys. Z., 19, 156
\bibitem{}Li, L. X., 2000, ApJ, 533, L115
\bibitem{}Li, L. X. 2002, ApJ, 567, 463
\bibitem{}Lovelace, R.V.E., Berk, H.L., Contopoulos, J. 1991, ApJ, 379, 696
\bibitem{}Lyutikov, M., \& Blandford R. 2003 [arXiv:astro-ph/0312347]
\bibitem{}MacDonald, D., {\&} Thorne, K. S. 1982, MNRAS, 198, 345
\bibitem{}MacFadyen, A. I., {\&} Woosley, S. E. 1999, ApJ, 524, 262
\bibitem{}Norris, J. P., et al. 1996, ApJ, 459, 393
\bibitem{}Novikov, I. D., {\&} Thorne, K. S. 1973, in Black Holes,
ed. Dewitt C (Gordon and Breach, New York) p.345
\bibitem{}Paczynski, B. 1986, ApJ, 308, L43
\bibitem{}Piran, T. 2004, Rev. Mod. Phys., 76, 1143
\bibitem{}Portegies Zwart, S. F., Lee, C. H., {\&} Lee, H. K. 1999, ApJ, 529,
666 (PZLL99)
\bibitem{}Portegies Zwart, S. F., {\&} Totani, T. 2001, ApJ, 328, 951 (PZT01)
\bibitem{}Reynoso, M. M., Romero, G. E., {\&} Sampayo, O. A. 2006, A{\&}A,
454, 11 (astro-ph/0511639) (RRS06)
\bibitem{}Romero, G. E., Torres, D. F., Andruchow, I., {\&} Anchordoqui, L. A.
1999, MNRAS, 308, 799
\bibitem{}Sakurai, T. 1985, A{\&}A, 152, 121
\bibitem{}Sari, R., Narayan, R., {\&} Piran, T. 1996, ApJ, 473, 204
\bibitem{}Spruit, H. C. 1994, in Cosmical Magnetism, ed. D.
Lynden-Bell, Dordrecht: Kluwer, p. 33
\bibitem{}Spruit, H. C., Daigne F., {\&} Drenkhahn G. 2001, A{\&}A, 369, 694
\bibitem{}Wang, D. X., Ma, R. Y., Lei, W. H., {\&} Yao, G. Z. 2003, ApJ, 595,
109
\bibitem{}Wang, D. X., Xiao, K., {\&} Lei, W. H. 2002, MNRAS, 335, 655
\bibitem{}Wang, D. X., Lei, W. H., {\&} Ye, Y. C. 2006, ApJ, 643, 1047
\bibitem{}Woosley, S. E. 1993, ApJ, 405, 273
\bibitem{}van Putten, M. H. P. M. 1999, Science, 284, 115
\bibitem{}van Putten, M. H. P. M. 2001, Phys. Rep., 345, 1
\bibitem{}van Putten, M. H. P. M., {\&} Ostriker, E. C. 2001, ApJ,
552, L31
\bibitem{}van Putten, M. H. P. M., {\&} Levinson, A. 2003, ApJ, 584,
937
\bibitem{}van Putten, M. H. P. M., Levinson, A., Regimbau, T.,
Punturo, M., {\&} Harry, G. M. 2004, Phys. Rev. D, 69, 044007

\end{thebibliography}
\end{document}